# Use of ARAS 360 to Facilitate Rapid Development of Articulated Total Body Biomechanical Physics Simulations

Bob J. Scurlock, Ph.D., ACTAR
*Department of Physics, University of Florida, Gainesville, Florida*

**Introduction**

The development of 3-dimensional environments to be used within a biomechanical physics simulation framework, such as Articulated Total Body, can be laborious and time intensive. This brief article demonstrates how the ARAS 360 software package can aid the user by speeding up development time.

**The Challenge of Articulated Total Body**

Articulated Total Body (ATB) is a general-purpose 3-dimensional physics simulation package typically used to simulate interactions of the human body with its environment, including automobiles. A brief introduction to ATB is given in Reference [1]. A human body can be approximated within ATB as a series of ellipsoids connected by joints with configurable properties. These ellipsoids are then made to interact with various contact planes according to the laws of physics.

One possible application with the ATB package involves pedestrian impacts. The difficulty one typically encounters with modeling a passenger vehicle impacting a pedestrian, however, is the geometrical definitions of the contact planes themselves. These contact planes need to geometrically approximate the surface of the vehicle exterior. The definition of the force responses of the contact planes to deflection and deflection rate is beyond the scope of this article. In ATB, the user must specify the three vertices to uniquely define every contact plane within the environment. These planes can approximate body panels on a vehicle or the roadway itself. How best to approximate the geometrical surface structure of a vehicle is a challenge, but the ARAS 360 software package makes the task tractable and fast [2].

**Using ARAS to Model Surfaces**

Suppose one needs to model an impact between a newer model Porsche 911 Carrera and an unfortunate jogger running across a busy intersection. Using ARAS 360, one can access the large database of vehicles and select the appropriate vehicle (Figure 1). ARAS provides excellent vehicle models with well-behaved vehicle meshes that are generally good approximations to the actual vehicles themselves.

After selecting the appropriate vehicle model in the ARAS environment, one can add a simple rectangle to the environment (Figure 2). Using the size and orientation controls within ARAS, the rectangle can be stretched, rotated, and modified until it reasonably approximates the surface of interest on the vehicle. In our example, we can first approximate the hood of the Porsche. This is shown in Figure 3, along with the definitions of the rectangle's "Height" ($H$) and "Length" ($L$).

The ATB package requires three vertices to be defined for each contact plane. Using the three vertices, two surface vectors can be constructed to fall within the contact plane, and whose cross-product defines the normal surface vector. With some basic trigonometry, the position of the three required vertices can be obtained from the information given by ARAS and converted into the format suitable for ATB (Figure 4).

Generally, the ATB coordinate system does not necessarily fall along the same orientation as in ARAS. For example, the $z$-axis points upward vertically in ARAS, whereas in ATB it generally points downward. The $y$-axis is inverted as well. To transform between ARAS space and ATB space for our purposes, one can simply perform the operation $(x,y,z)_{ARAS} \rightarrow (x, -y, -z)_{ATB}$ for any given point. This is easily accommodated in the spreadsheet. The vertex calculations can be generalized such that after a round of debugging, in which the user verifies the proper orientation of one plane in ATB, additional contact planes can be rapidly developed. We will return to this topic at the end.

Here, a mathematically elegant approach is developed for the case of contact planes that have mirror symmetry about the $x$-$z$ plane. In the case of our rectangular plane approximating the hood, we must first identify the location of three vertices that uniquely characterize the plane's geometry. Let us identify the three surface vertices as $\bar{R}_1, \bar{R}_2, \bar{R}_3$, where these vectors are defined with respect to the vehicle coordinate frame. Here, the origin is given by ($x_{CG}, y_{CG}, 0$). $x_{CG}$ and $y_{CG}$ are the $x$- and $y$- coordinates of the vehicle center-of-gravity. The $z$-coordinate is taken at ground level. Let us denote the vector $(x',y',z')$ as the position of the contact plane's geometrical center (see Figure 4). The plane's geometrical center can be obtained by first selecting the snap to point feature in ARAS. Draw a line within the environment, and set its start point to (0,0,0). If one drags the end point of the line near the center of the contact plane, the snap to point feature should automatically set the end point of the line at the plane's geometrical center. All one has to do then is note the $(x,y,z)$ position of the end point in the line properties dialogue to the left. This is depicted in Figure 5.

Let us call the angle made between the $x$-$y$ plane and the contact surface plane the pitch, which we will denote by $\theta$. The pitch can also be understood as the angle made between the contact plane surface normal vector and the $z$-axis. We can write the following expressions for the locations of the vertices in the vehicle frame:

$$\bar{R}_1 = \begin{pmatrix} x_1 \\ y_1 \\ z_1 \end{pmatrix} = \begin{pmatrix} x' - \frac{L}{2}\cos\theta \\ y' - \frac{H}{2} \\ z' + \frac{L}{2}\sin\theta \end{pmatrix}$$

$$\bar{R}_2 = \begin{pmatrix} x_2 \\ y_2 \\ z_2 \end{pmatrix} = \begin{pmatrix} x' + \frac{L}{2}\cos\theta \\ y' - \frac{H}{2} \\ z' - \frac{L}{2}\sin\theta \end{pmatrix}$$

$$\bar{R}_3 = \begin{pmatrix} x_3 \\ y_3 \\ z_3 \end{pmatrix} = \begin{pmatrix} x' - \frac{L}{2}\cos\theta \\ y' + \frac{H}{2} \\ z' + \frac{L}{2}\sin\theta \end{pmatrix}$$

Sitting in the driver's seat, the vector describing the right side of the plane is given by:

$$\bar{d}_{21} = \bar{R}_2 - \bar{R}_1 = \begin{pmatrix} x' + \frac{L}{2}\cos\theta - \left[x' - \frac{L}{2}\cos\theta\right] \\ y' - \frac{H}{2} - \left[y' - \frac{H}{2}\right] \\ z' - \frac{L}{2}\sin\theta - \left[z' + \frac{L}{2}\sin\theta\right] \end{pmatrix}$$

$$= \begin{pmatrix} L\cos\theta \\ 0 \\ -L\sin\theta \end{pmatrix}$$

As expected, we have $\|\bar{d}_{21}\| = L$. The vector describing the side of the plane nearest to the vehicle center-of-gravity is given by (see Figure 4):

$$\bar{d}_{31} = \bar{R}_3 - \bar{R}_1 = \begin{pmatrix} x' - \frac{L}{2}\cos\theta - \left[x' - \frac{L}{2}\cos\theta\right] \\ y' + \frac{H}{2} - \left[y' - \frac{H}{2}\right] \\ z' + \frac{L}{2}\sin\theta - \left[z' + \frac{L}{2}\sin\theta\right] \end{pmatrix}$$

$$= \begin{pmatrix} 0 \\ H \\ 0 \end{pmatrix}$$

Here we have $\|\bar{d}_{31}\| = H$.

Finally, we can take the normalized cross-product $\bar{d}_{21} \times \bar{d}_{31}$ to verify the normal vector to the contact surface. This is given by:

$$\frac{\bar{d}_{21} \times \bar{d}_{31}}{\|\bar{d}_{21} \times \bar{d}_{31}\|} = \begin{vmatrix} \hat{x} & \hat{y} & \hat{z} \\ L\cos\theta & 0 & -L\sin\theta \\ 0 & H & 0 \end{vmatrix}$$

$$= \begin{pmatrix} \sin\theta \\ 0 \\ \cos\theta \end{pmatrix} = \hat{n}$$

This is indeed the unit normal surface vector for the contact surface. This vector is shown in Figure 4. In ATB, the direction of this normal surface vector describes the direction that the contact force will take within physics simulation.

In our simple example, a total of five contact planes were defined to characterize all of the surfaces of the Porsche with which the ATB-simulated human might interact. These planes are shown in Figure 6. Since the ATB package requires contact plane vertices to be defined in a particular way, it is best to use a simple spreadsheet tool in which one can insert the geometrical properties and orientation of the rectangular surfaces from ARAS, such as the Length, Height, Pitch, Roll, Yaw, and the (*x'*,*y'*,*z'*) position of the plane-centers within the ARAS environment to obtain the required vertex locations. Figure 7 shows the output of such a spreadsheet tool. The spreadsheet was programmed with the equations shown above. Figure 8 depicts the contact planes as visualized in the ARAS and ATB environments.

Finally, the impact sequence can be seen in Figure 9, and a video of the final, rendered animation can been viewed online [3]. While the topic of compositing with ARAS is left for a future article, the raw ATB animation output was merged with a pre-impact sequence generated with ARAS to form the final output, which can also be viewed online [4]. This final composite is depicted in Figure 10.

**An Even Faster Method**

As stated previously, the approach outlined above will work in general for mirror symmetric planes about the *x-z* plane, and the benefit of its mathematical development presented above allows for quick verification of the contact force directions simulated within ATB; however, the equations will need to be modified for planes rotated in yaw and/or roll, where *x-z* plane symmetry is spoiled. As a general – and likely faster – approach, one can always simply sample the positions of each vertex explicitly in ARAS as was illustrated above to obtain the position of the geometric center of each rectangle. Again, to perform this task, one simply draws a line object in the scene, locking the start point to the origin using the snap to point option. One then drags the end point to each vertex and logs the position given in the properties dialogue on the left. The three lines needed to obtain the three vertices for our hood contact plane are shown in Figure 11.

**Conclusion**

In this article, a simple technique has been demonstrated to quickly develop an ATB-based pedestrian impact simulation using the powerful features of the ARAS 360 software package. The total development time for the example shown here was on the order of a few hours. The same technique can also be used to model the interior space of the occupant cabin, as features of the cabin are also included in the ARAS vehicle models. By importing those potential contact planes into ATB, one could perform standard occupant kinematic simulations.

*Bob Scurlock, Ph. D. is a Research Associate at the University of Florida, Department of Physics and works as a consultant for the accident reconstruction and legal community. He can be reached at BobScurlockPhD@gmail.com. His website offering free analysis software can be found at: ScurlockPhD.com.*

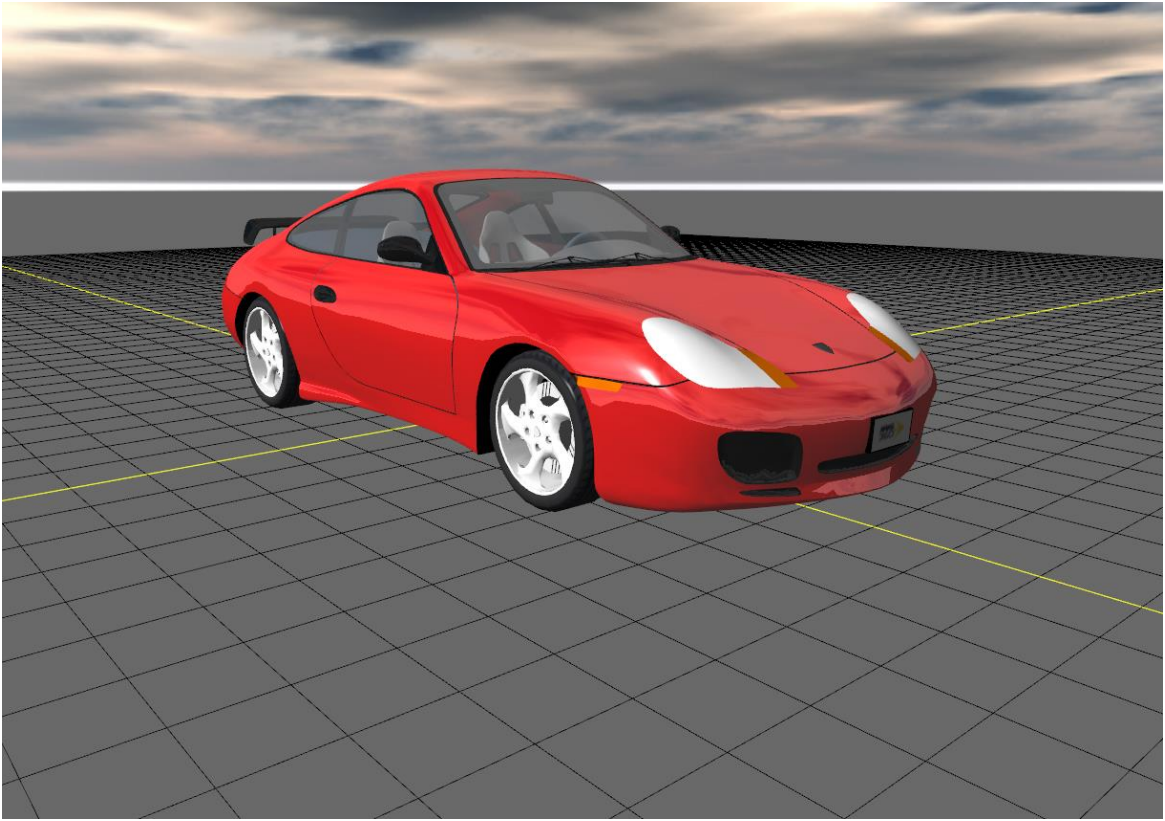

**Figure 1: Porsche 911 Carrera selected from vehicle database.**

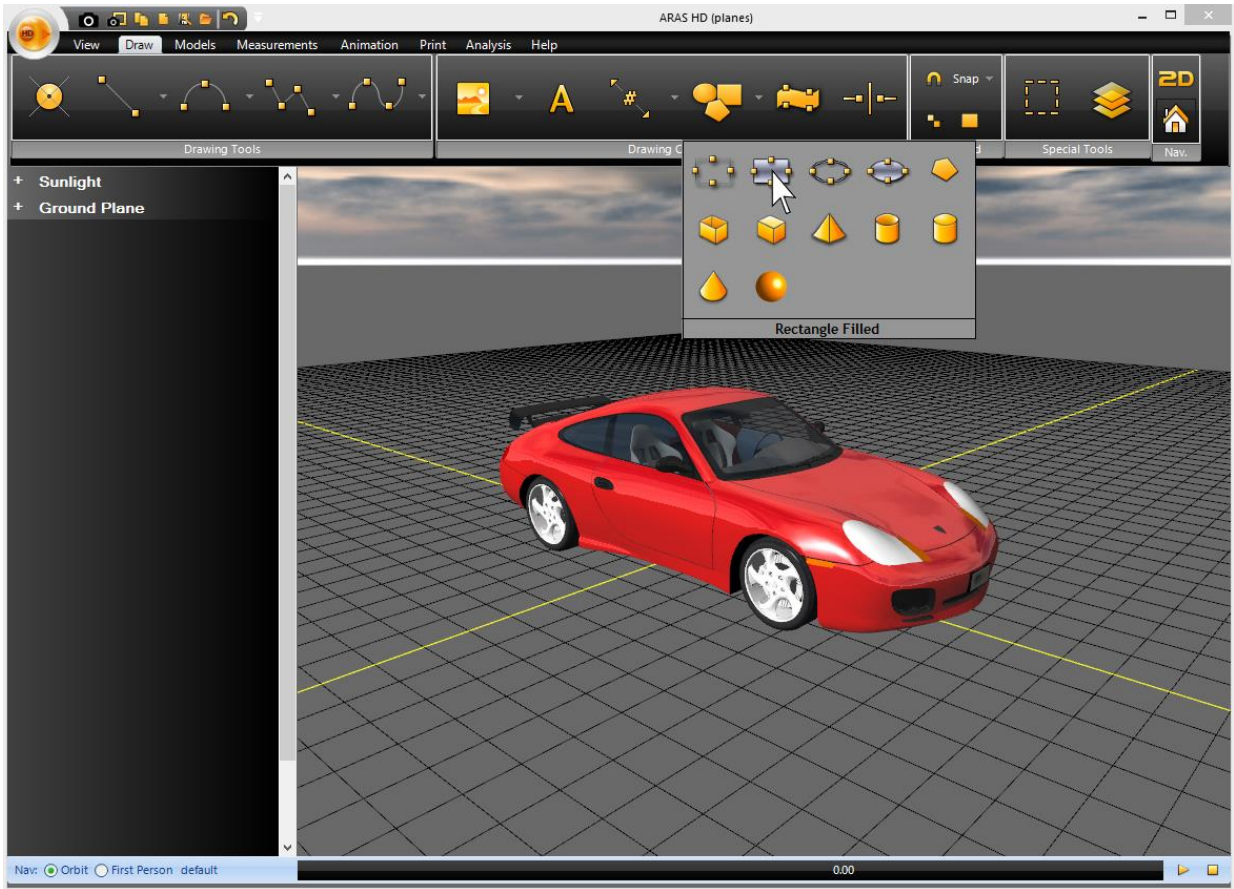

Figure 2: Selecting the draw rectangle tool.

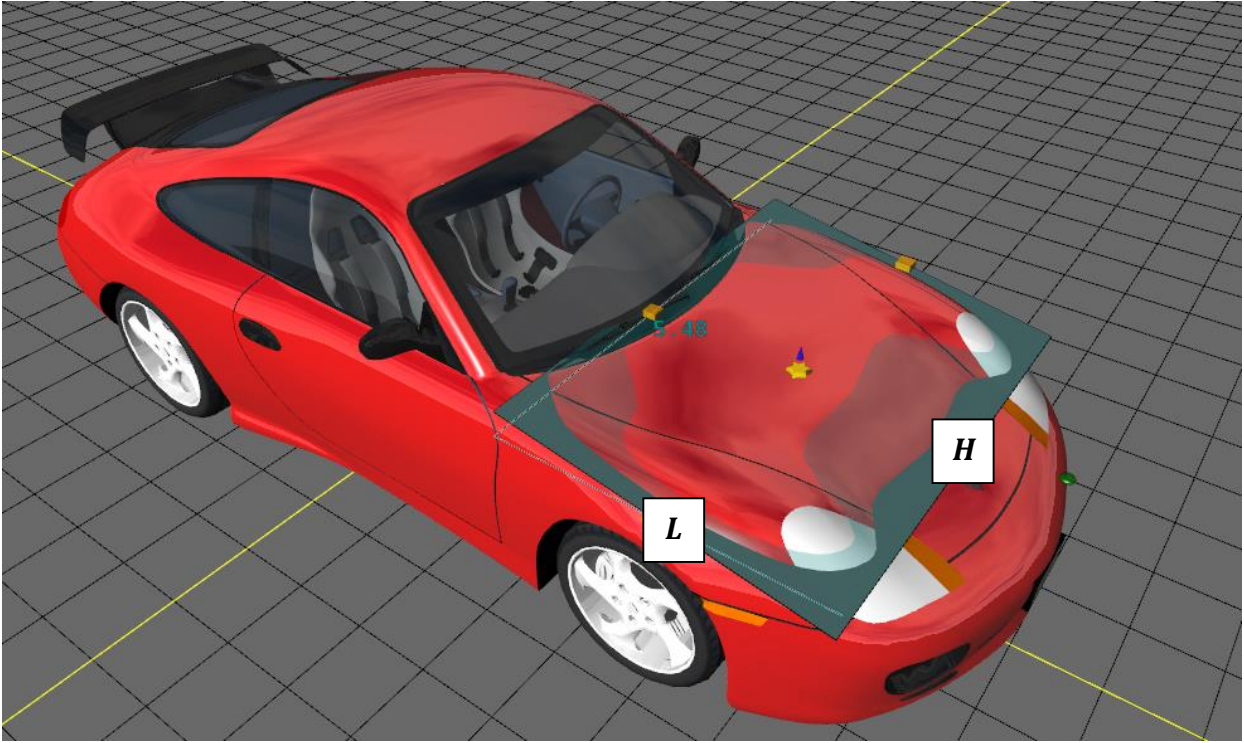

**Figure 3: Rectangle is moved and scaled so as to approximate the hood. The sides corresponding to "Length" (*L*) and "Height" (*H*) are shown.**

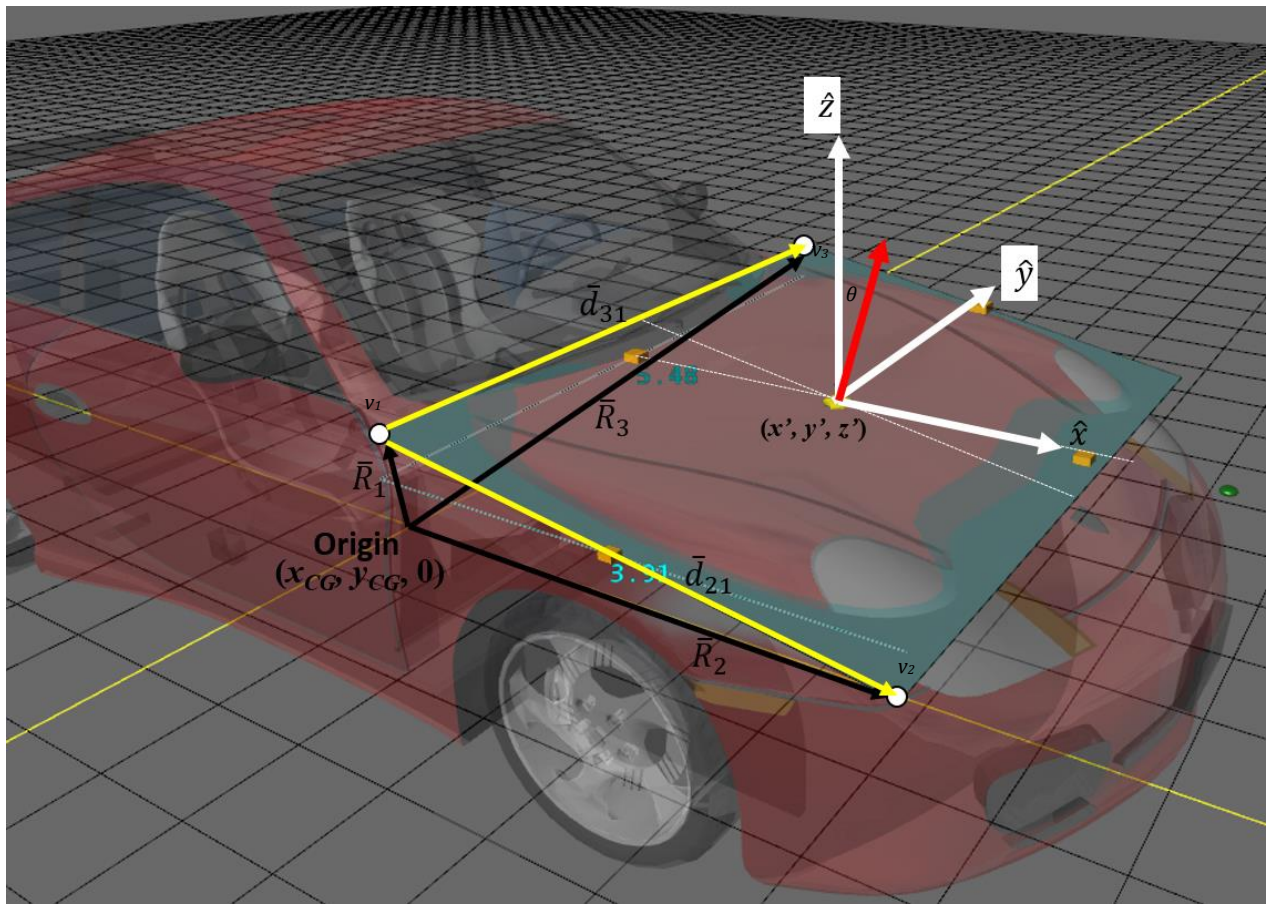

**Figure 4: Surface geometry vectors shown for hood contact plane. Vertices $v_1$, $v_2$, and $v_3$ are shown as well as the vectors $R_1$, $R_2$, and $R_3$ defining their positions in the vehicle reference frame.**

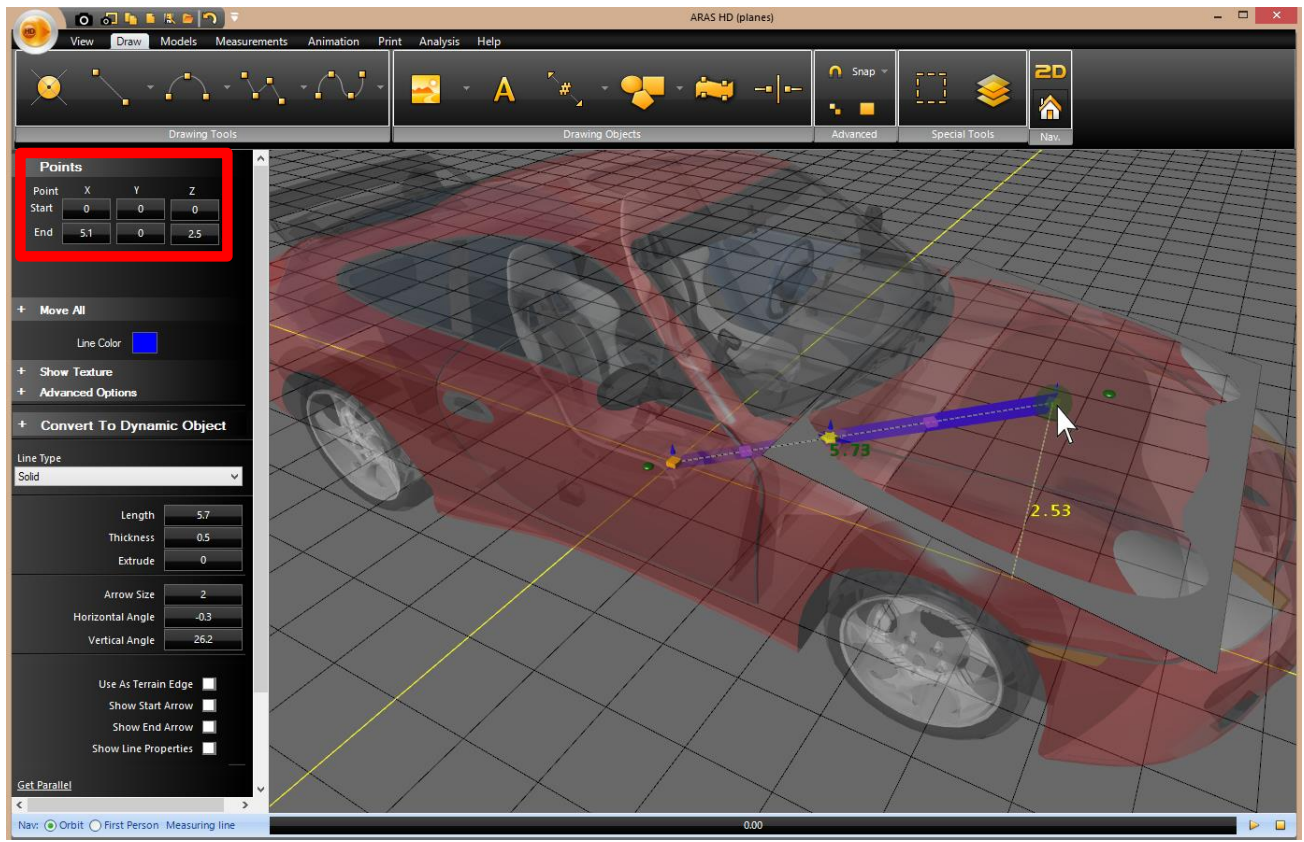

Figure 5: Method of obtaining geometrical center of contact plane. A line is drawn from the origin to the center of the plane. Using the snap to point feature in ARAS, the line end point will automatically go to the center of the contact plane. The end point location can then be noted in the line properties dialogue shown in the red square.

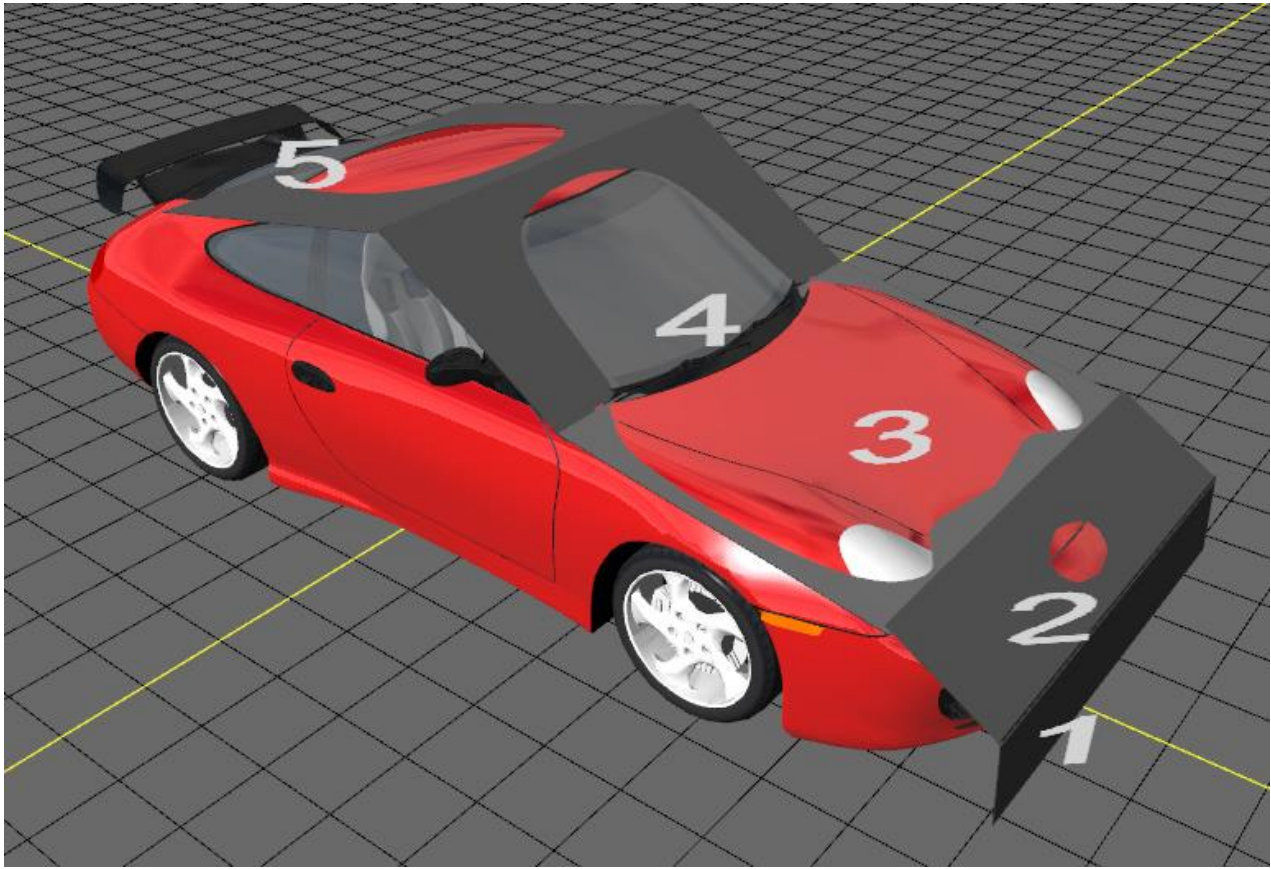

**Figure 6: Five contact planes used to characterize geometry of vehicle.**

| | A | B | C | D | E | F | G | H | I | J | K | L | M | N | O | P | Q | R |
|---|---|---|---|---|---|---|---|---|---|---|---|---|---|---|---|---|---|---|
| 1 | ARAS Planes | x | y | z | Length | Height | Pitch | Yaw | Roll | x1 | y1 | z1 | x2 | y2 | z2 | x3 | y3 | z3 |
| 2 | 1 | 8.20 | 0.00 | 1.10 | 1.00 | 5.48 | 86.00 | 0.00 | 0.00 | 8.17 | -2.74 | 1.60 | 8.23 | -2.74 | 0.60 | 8.17 | 2.74 | 1.60 |
| 3 | 2 | 7.60 | 0.00 | 1.90 | 1.42 | 5.48 | 31.00 | 0.00 | 0.00 | 6.99 | -2.74 | 2.27 | 8.21 | -2.74 | 1.53 | 6.99 | 2.74 | 2.27 |
| 4 | 3 | 5.10 | 0.00 | 2.50 | 3.91 | 5.48 | 10.00 | 0.00 | 0.00 | 3.17 | -2.74 | 2.84 | 7.03 | -2.74 | 2.16 | 3.17 | 2.74 | 2.84 |
| 5 | 4 | 2.00 | 0.00 | 3.60 | 3.00 | 5.48 | 33.00 | 0.00 | 0.00 | 0.74 | -2.74 | 4.42 | 3.26 | -2.74 | 2.78 | 0.74 | 2.74 | 4.42 |
| 6 | 5 | -1.20 | 0.00 | 4.00 | 3.91 | 5.48 | -11.00 | 0.00 | 0.00 | -3.12 | -2.74 | 3.63 | 0.72 | -2.74 | 4.37 | -3.12 | 2.74 | 3.63 |
| 7 | | | | | | | | | | | | | | | | | | |
| 8 | ATB Planes | | | | | | | | | x1 | y1 | z1 | x2 | y2 | z2 | x3 | y3 | z3 |
| 9 | 1 | | | | | | | | | 98.0 | 32.9 | -19.2 | 98.8 | 32.9 | -7.2 | 98.0 | -32.9 | -19.2 |
| 10 | 2 | | | | | | | | | 83.9 | 32.9 | -27.2 | 98.5 | 32.9 | -18.4 | 83.9 | -32.9 | -27.2 |
| 11 | 3 | | | | | | | | | 38.1 | 32.9 | -34.1 | 84.3 | 32.9 | -25.9 | 38.1 | -32.9 | -34.1 |
| 12 | 4 | | | | | | | | | 8.9 | 32.9 | -53.0 | 39.1 | 32.9 | -33.4 | 8.9 | -32.9 | -53.0 |
| 13 | 5 | | | | | | | | | -37.4 | 32.9 | -43.5 | 8.6 | 32.9 | -52.5 | -37.4 | -32.9 | -43.5 |

**Figure 7: Simple spreadsheet tool to calculate vertex locations and transform them to ATB space.**

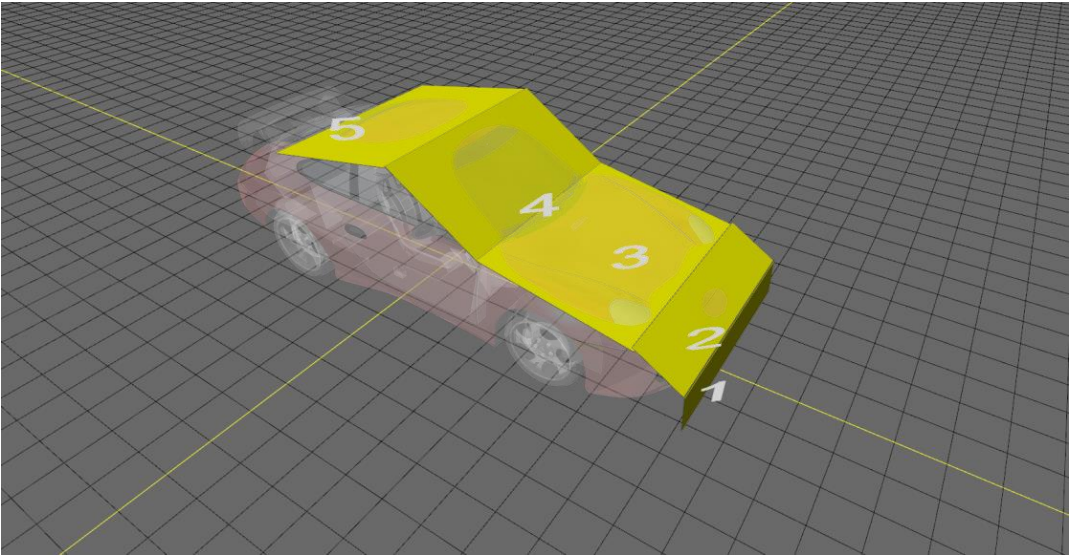

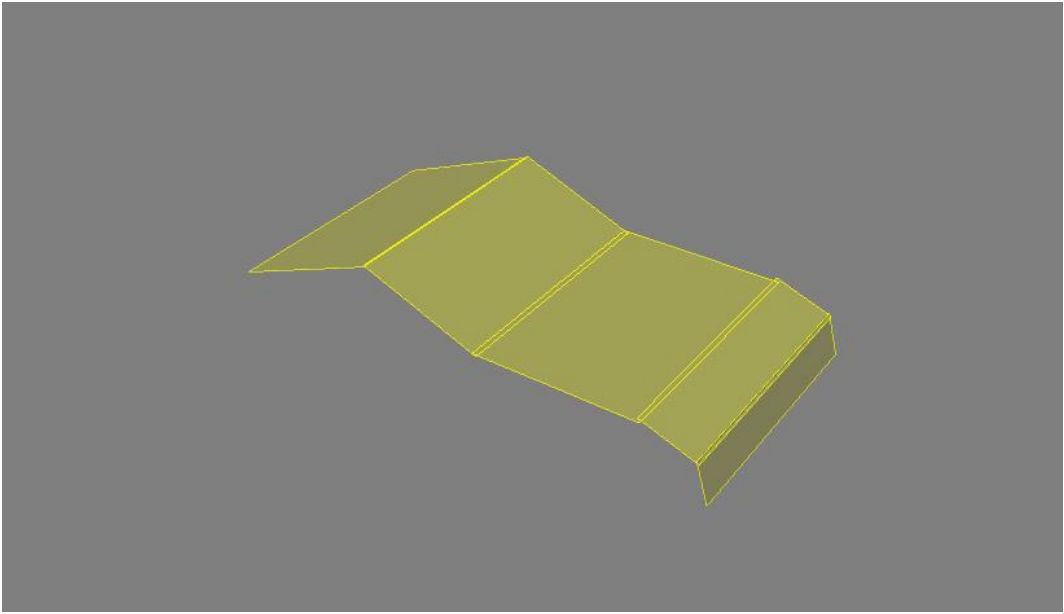

**Figure 8: Top: Contact planes in ARAS 360 HD environment. Bottom: Contact planes shown within the ATB environment.**

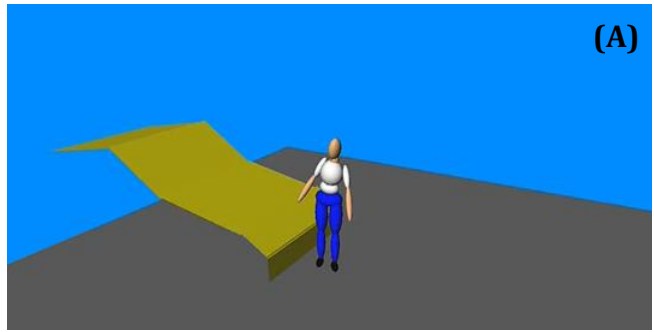

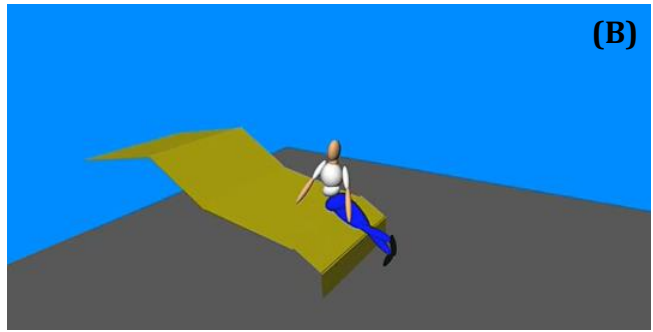

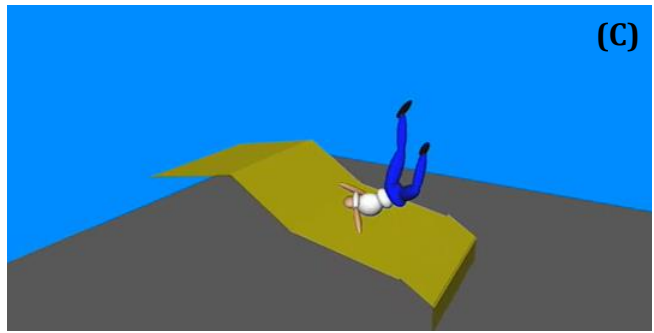

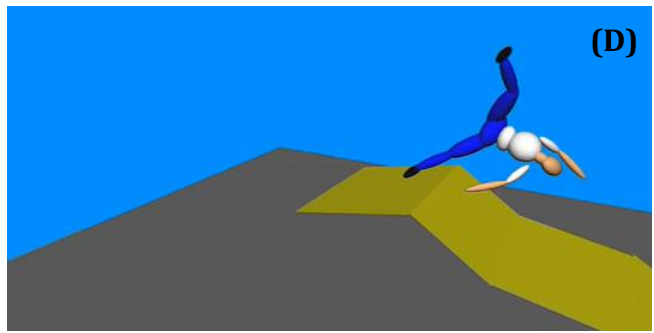

**Figure 9: Impact sequence visualized within ATB.**

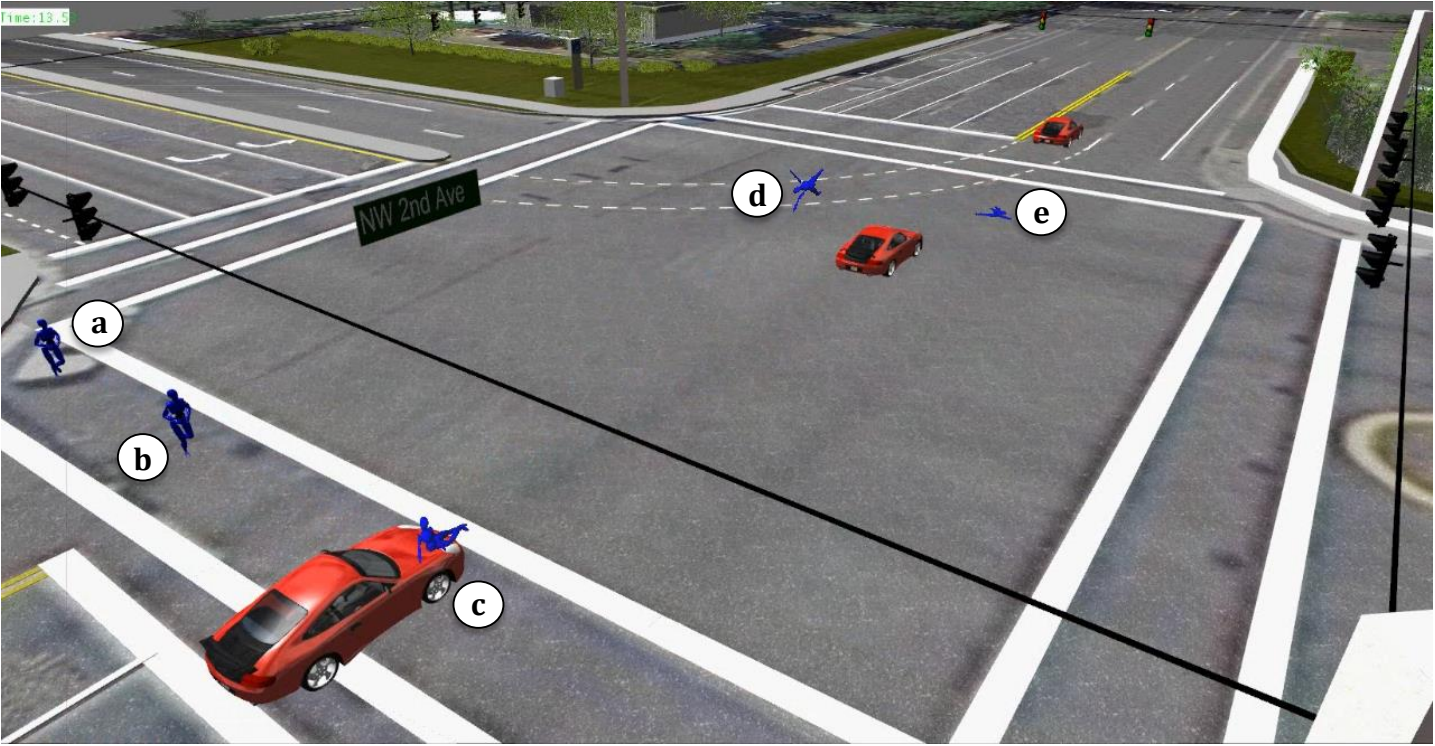

**Figure 10: Animation sequence from ARAS and ATB composite video.**

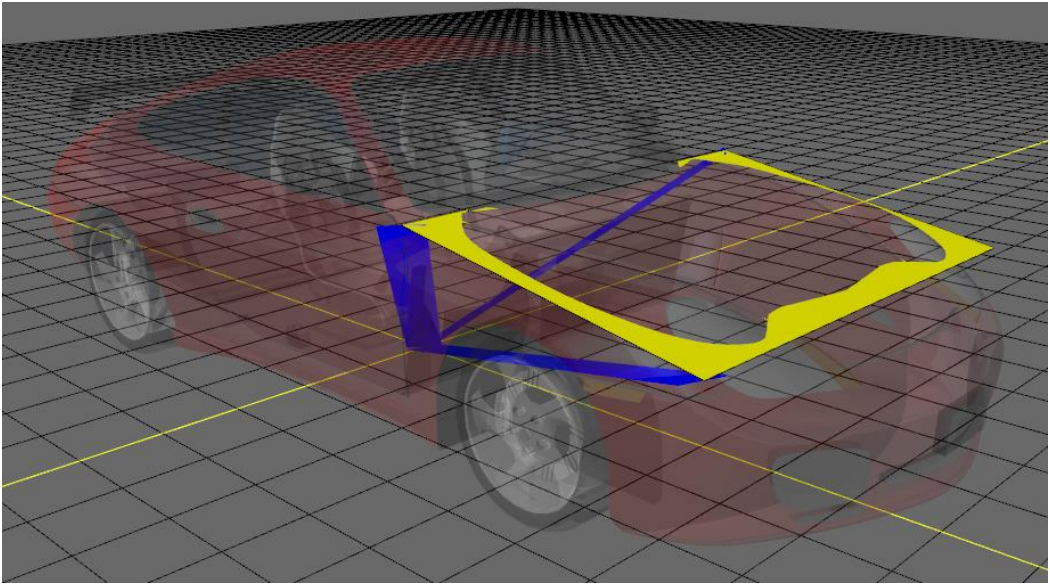

**Figure 11: Explicit method to obtain vertex positions. Lines drawn from origin to vertices are shown.**